\documentclass{ws-procs9x6}
\usepackage{epsfig}

\begin{document}

\title{
\vspace*{-41pt}
STUDY OF CHARGE MULTIPLICITY IN HADRONIC THREE-JET Z DECAYS AT LEP
\footnote{\uppercase{T}alk given at the 32nd \uppercase{I}nternational 
\uppercase{S}ymposium on \uppercase{M}ultiparticle \uppercase{D}ynamics, 
\uppercase{A}lushta, \uppercase{U}kraine, \uppercase{S}eptember 7\,--\,13, 
2002.}}

\author{Vladimir~A.~UVAROV}

\address{Institute for High Energy Physics, Protvino, Russia \\
(for the DELPHI Collaboration)}

\maketitle

\vspace*{-8pt}
\abstracts{
The mean charge multiplicity in hadronic three-jet Z decays has been measured 
with the DELPHI detector as a function of the event topology and compared with
recent theoretical calculations. 
The QCD colour factor ratio $C_A/C_F$ was determined, and the gluon 
contribution to this multiplicity was extracted giving a measurement of the 
mean charge multiplicity of a two-gluon colour-singlet system as a function 
of the effective c.m.~energy covering the range from 16 to 52 GeV.}

\vspace*{-8pt}
The radiation of gluons by quarks (other gluons) is proportional to the 
colour charge $C_F$ ($C_A$), where $C_F$ and $C_A$ are the eigenvalues of 
the quadratic Casimir-operators of the $SU(3)_C$ group. 
Therefore, the comparison of the contributions of quark and gluon jets to the 
mean particle multiplicity in hadronic Z decays gives insight into the group 
structure of the QCD. 
These contributions can be studied within the framework of the colour dipole
model\,\cite{Eden:1998ig,Eden:1999vc} which predicts the mean charge 
multiplicity of hadronic three-jet $e^+e^-$ events as a function of the event 
topology.
Recently this model has been applied\,\cite{note1,Abbiendi:2001us} to 
investigate the multiplicity of {\em symmetric} ${\rm Z} \to q\bar{q}g$ 
events only with one relevant scale.
To provide a more thorough test of this model, in the present 
study,\cite{ABS247} results of which are preliminary, the multiplicity 
is measured for {\em asymmetric} ${\rm Z} \to q\bar{q}g$ events as a function 
of two relevant scales, 
then the colour factor ratio $C_A/C_F$ and the gluon contribution to this 
multiplicity are determined.

The data used in this analysis have been collected with the DELPHI detector
at $\sqrt{s}=M_Z$ in the years 1992--95. 
The reconstruction and selection criteria of charged and neutral particles 
and hadronic events have been described in Ref.\cite{Abreu:1999rs}. 
All selected particles in each selected event were clustered to a three-jet 
configuration using the Cambridge,\cite{Dokshitzer:1997in} the angular 
ordered Durham\,\cite{Dokshitzer:1997in} and the Durham\,\cite{Catani:1991hj} 
algorithms without a specified value of the jet resolution parameter 
$y_{cut}$.
The correction factor for the detector acceptance and inefficiencies 
calculated from the Monte Carlo simulation as a function of the event 
topology was applied to the data. 

The topology of the event with three massless jets can be described by two 
of three inter-jet angles numbered in order of size 
($\theta_1 < \theta_2 < \theta_3$).
As $y_{cut}$ is adjusted separately for each event, only that 
($\theta_3,\theta_2$) range is used in the analysis where the mean value 
of $\log_{10}(y_2/y_3)$ is greater than 1 for each data point (here $y_n$ 
is the smallest test variable $y_{ij}$ in an event at the stage of clustering 
when $n+1$ jets are still present). 
Also the ($\theta_3,\theta_2$) range where the correction factor 
deviates from a smooth behaviour or the contribution to $\chi^2$ of the 
fit is too large is excluded from the analysis.
Finally, the chosen intervals are the following:
$120^\circ < \theta_3 < 155^\circ$ and $108^\circ < \theta_2 < 155^\circ$.

The colour dipole model\,\cite{Eden:1998ig} predicts the relationship between 
the scale evolution of the mean charge multiplicities of two-gluon, 
$N_{gg}(L)$, and quark-antiquark, $N_{q\bar{q}}(L)$, colour-singlet systems:

\smallskip
\smallskip
$\frac{d}{dL}N_{gg}(L+c_g-c_q) ~=~ 
(C_A/C_F)\cdot\frac{d}{dL}N_{q\bar{q}}(L)\cdot(1-\alpha_0\,c_r/L)
$\,, \hfill (1)

\smallskip
\smallskip
\noindent
where $L=\ln(s/\Lambda^2)$, 
$s=E^2_{c.m.}$, $\Lambda$ = 250 MeV, $c_g=11/6$, $c_q=3/2$, 
$c_r=10\,\pi^2/27-3/2$, $\alpha_0=6\,C_A/(11C_A-2N_f)$ and $N_f=5$.
The constant of integration of this differential equation has been fixed by 
the CLEO measurement\,\cite{Alam:1992ir} of the mean charge multiplicity in 
$\chi^{\,\prime}_{b}(J$=$2) \to gg$ decays.

The mean charge multiplicity, $N_{q\bar{q}g}$, of hadronic three-jet $e^+e^-$ 
events defined by a $k_{\perp}$-based jet finder without a specified 
resolution scale is given by two alternative expressions\,\cite{Eden:1999vc} 
with a single scale $k_{\perp}$ for the gluon jet: 

\smallskip
\smallskip
$N_{q\bar{q}g} ~=~ N_{q\bar{q}}(L_{q\bar{q}},k_{\perp Lu}) +
\frac{1}{2}\,N_{gg}(k_{\perp Le})$\,, \hfill (2a)   

\smallskip
$N_{q\bar{q}g} ~=~ N_{q\bar{q}}(L,k_{\perp Lu}) +
\frac{1}{2}\,N_{gg}(k_{\perp Lu})$\,, \hfill (2b)

\smallskip
\smallskip
\noindent
where $L_{q\bar{q}}=\ln(s_{q\bar{q}}/\Lambda^2)$,\, 
$k_{\perp Lu}=\ln(s_{qg}s_{g\bar{q}}/s\Lambda^2)$,\, 
$k_{\perp Le}=\ln(s_{qg}s_{g\bar{q}}/s_{q\bar{q}}\Lambda^2)$,\, 
$s_{ij}=(p_i+p_j)^2$. 
The term $N_{q\bar{q}}(L,k_{\perp})$ takes into account that the resolution 
of the gluon jet at a given $k_{\perp}$ implies restrictions on the phase 
space of the $q\bar{q}$ system. 
It can be obtained from $N_{q\bar{q}}(L)$ through the following 
relation:\cite{Eden:1998ig}

\smallskip
\smallskip
$N_{q\bar{q}}(L,k_{\perp}) ~=~ N_{q\bar{q}}(L'\,) + 
(L-L'\,)\cdot\frac{d}{dL'}N_{q\bar{q}}(L'\,)$ ~with~ $L'=k_{\perp}+c_q$.
\hfill (3)

\smallskip
\smallskip
As for massless jets all dynamical scales are given by the inter-jet angles, 
Eqs.~(1), (2a,b) and (3) can be used to fit the ratio $C_A/C_F$ to the data. 
$N_{q\bar{q}}(L)$ is equivalent to the mean charge multiplicity of hadronic 
$e^+e^-$ events as a function of the c.m.~energy (measured by 
experiments\,\cite{qqbar} and corrected not to include the contribution of 
$b\bar{b}$ events). 
As in this analysis no $b$-tagging procedure is applied to the data,
an additive offset parameter $N_0$ is introduced to the fitting procedure
($N_{q\bar{q}g}=N_{q\bar{q}g}^{exp}-N_0$) to account for the extra
multiplicity due to the decay of hadrons containing the $b$ quark.
The value of this parameter is expected to be of the order of 
$N_0\simeq0.62$\,\cite{Akers:1995ww,Abreu:1995pk} and constant over a wide 
energy range.\cite{Akers:1995ww} 

\begin{figure}[tb]
\begin{center}
\mbox{\epsfig{file=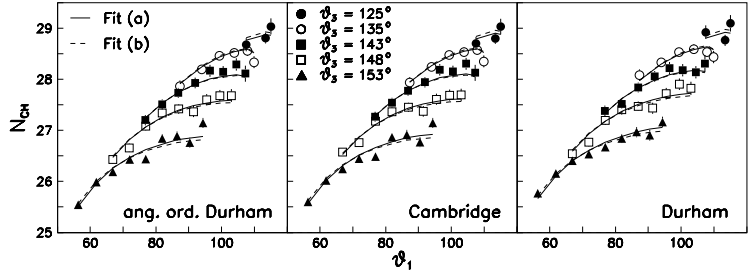,width=0.87\textwidth}}
\end{center}
\vspace*{-11pt}
\caption{Mean charge multiplicity $N_{ch}$ in hadronic three-jet Z decays 
as a function of the event topology ($\theta_1,\theta_3$) for three clustering 
algorithms. The curves are the results of the fits for two alternative 
predictions of Eqs.~(2a) and (2b). See text. \label{f_cafit}}
\vspace*{-13pt}
\end{figure}
\begin{table}[tb]
\tbl{The results of the fits for both predictions and all clustering 
algorithms.\vspace*{1pt}}
{\footnotesize
\begin{tabular}{|c|l|c|c|c|}
\hline
pred. & jet~finder & $C_A/C_F$ & $N_0$ & $\chi^2/ndf$ \\
\hline
(2a) & Cambridge        & 2.300~$\pm$~0.01 & 0.69~$\pm$~0.03 & 0.92 \\
(2a) & ang. ord. Durham & 2.329~$\pm$~0.03 & 0.60~$\pm$~0.08 & 1.03 \\
(2a) & Durham           & 2.228~$\pm$~0.02 & 0.97~$\pm$~0.03 & 0.89 \\ 
\hline
(2b) & Cambridge        & 2.11~~$\pm$~0.06 & 0.36~$\pm$~0.08 & 1.17 \\
(2b) & ang. ord. Durham & 2.151~$\pm$~0.04 & 0.27~$\pm$~0.04 & 1.33 \\
(2b) & Durham           & 2.000~$\pm$~0.05 & 0.60~$\pm$~0.06 & 0.96 \\
\hline
\end{tabular}\label{t_cafitres} }
\vspace*{-13pt}
\end{table}

The gluon jet is not identified explicitly in this analysis. 
Therefore, the theoretical predictions for $N_{q\bar{q}g}$ at a given topology 
($\theta_3,\theta_2$) with the gluon jet being {\em jet\,1}, {\em jet\,2} or 
{\em jet\,3} are added with proper weights obtained from the QCD three-jet 
matrix element in the first order. 
Furthermore, the hadronization process affects the inter-jet angles by 
pulling close jets. 
Therefore, the corresponding correction calculated from the Ariadne 
Monte Carlo simulation by comparing the multiplicities defined at the 
partonic level with the ones ``measured'' at the hadronic level
is taken into account.

Fig.~\ref{f_cafit} shows the {\em measured} mean charge multiplicities, 
$N_{ch}$, in hadronic three-jet Z decays as a function of $\theta_1$ for 
five $\theta_3$ intervals and three clustering algorithms. 
The results of the fits for both predictions and all clustering algorithms 
(see Table~\ref{t_cafitres}) agree well with the data shown in 
Fig.~\ref{f_cafit}. 
The values of parameters fitted with Eq.~(2a) are higher than those fitted 
with Eq.~(2b) and mostly agree better with the expectation of $C_A/C_F=9/4$ 
and $N_0\simeq0.62$. 
A weighted average of the values obtained for all clustering algorithms 
results in a value of $C_A/C_F=2.277 \pm 0.02 \pm 0.05$ for Eq.~(2a) and 
$C_A/C_F=2.093 \pm 0.05 \pm 0.08$ for Eq.~(2b) (the first error is statistical
and the second one is systematic).
One can see that the first value is more consistent with the QCD expectation. 

\begin{figure}[tb]
\begin{center}
\mbox{\epsfig{file=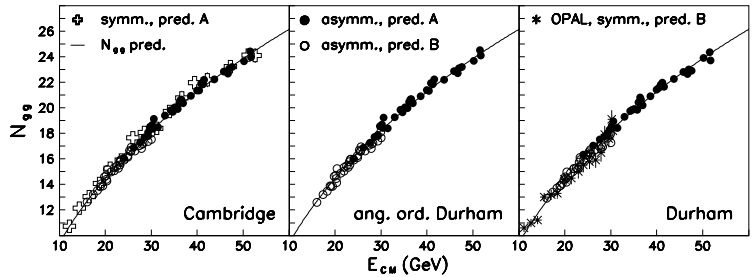,width=0.87\textwidth}}
\end{center}
\vspace*{-11pt}
\caption{Mean charge multiplicity $N_{gg}(k_{\perp})$ of a two-gluon 
colour-singlet system as a function of the effective c.m.~energy $E_{c.m.}$ 
given by $k_{\perp Le}$ for Eq.~(2a) and $k_{\perp Lu}$ for Eq.~(2b). 
The curves show the prediction of Eq.~(1). See text. \label{f_gluons}}
\vspace*{-9pt}
\end{figure}
Eqs.~(2a,b) can be used to extract the multiplicity $N_{gg}(k_{\perp})$ from 
the measured values of $N_{ch}(\theta_1,\theta_3)$ shown in 
Fig.~\ref{f_cafit}.
The results of this extraction (with fitted $N_0$ and fixed $C_A/C_F=9/4$) 
are shown in Fig.~\ref{f_gluons} for both predictions and all clustering 
algorithms over the effective c.m.~energy range from 16 to 52 GeV.
These results agree well with the prediction of Eq.~(1) and with the previous 
DELPHI\,\cite{note1} and OPAL\,\cite{Abbiendi:2001us} results obtained only
for symmetric three-jet ${\rm Z} \to q\bar{q}g$ decays.

In conclusion, the mean charge multiplicity in hadronic three-jet Z decays 
has been measured as a function of the event topology ($\theta_1,\theta_3$).
Fits of the two theoretical predictions to these data yield two values for the 
colour factor ratio $C_A/C_F$ of $2.277 \pm 0.02 \pm 0.05$ and 
$2.093 \pm 0.05 \pm 0.08$. 
Also the mean charge multiplicity of a two-gluon colour-singlet system 
has been extracted as a function of the effective c.m.~energy, in good 
agreement with previous measurements for symmetric three-jet Z decays.

\end{document}